\documentclass{PoS}

\title{A test of first order scaling in $N_f$=2 QCD}

\ShortTitle{A test of first order scaling in $N_f$=2 QCD}

\author{Guido Cossu\\
SNS and INFN Pisa, Piazza dei Cavalieri 7, I-56127 Pisa, ITALY\\
\email{g.cossu@sns.it}}

\author{Massimo D'Elia\\
	Dipartimento di Fisica and INFN Genova, Via Dodecaneso 33, I-16146 Genova, ITALY\\
\email{delia@ge.infn.it}}

\author{Adriano Di Giacomo\\
	Dipartimento di Fisica and INFN Pisa, Largo Pontecorvo 3, I-56127 Pisa, ITALY\\
\email{digiaco@df.unipi.it}}

\author{\speaker{Claudio Pica}\\
	Brookhaven National Laboratory, Physics Department, Upton, NY 11973-5000, USA\\
\email{pica@bnl.gov}}

\abstract{
We complete our analysis of $N_f$=2 QCD based on the lattice
staggered fermion formulation. Using a series of Monte Carlo
simulations at fixed $m_L L_s^{y_h}$ one is able to test the
universality class with given critical exponent $y_h$. This strategy
has been used to test the $O(4)$ universality class and it has been
presented at the previous Lattice conferences. No agreement was found
with simulations in the mass range $am_q=[0.01335,0.15]$ using
lattices with $L_s=16$ up to $32$ and $L_t=4$.
With the same strategy, we now investigate the possibility of a first
order transition using a new set of Monte Carlo data corresponding to
$y_h=3$ in the same mass and volume range as the one used for $O(4)$.
A substantial agreement is observed both in the specific heat scaling
and in the scaling of the chiral condensate, while the chiral
susceptibilities still presents visible deviation from scaling in the
mass range explored.}

\FullConference{The XXV International Symposium on Lattice Field Theory\\
		 July 30-4 August 2007\\
		 Regensburg, Germany}

\begin{document}

\section{Motivation}

The order of the chiral transition of two-flavor QCD is still debated.
Its relevance is both phenomenological (heavy ion experiments) 
and theoretical: is deconfinement an order-disorder transition 
associated with a change of symmetry?

With massless quarks the chiral/de\-con\-fi\-ne\-ment transition for $N_f$=2 is 
expected to be second order in the $O(4)$ universality class or first order.
Two completely different scenarios correspond to those two
possibilities. Since, contrary to the first order case, 
second order phase transitions are
unstable against the explicit breaking of the underlying symmetry,
in the second case one would have a crossover instead of a real
phase transition for small but non-zero quark masses. That would 
imply the possibility of going continuously from a confined to a 
deconfined state of matter, in contrast
with the idea of confinement being an absolute property
of strongly interacting matter at zero temperature and of
deconfinement being an order-disorder transition associated 
to a change of symmetry~\cite{hooft}.
A second consequence would be the presence of a crossover line, at finite
mass, in the region of high temperature and small baryon chemical
potential $\mu_B$ of the QCD phase diagram, thus implying a critical
point~\cite{Stephanov} to connect with the first order line which is
supposed to exist at low temperatures and large chemical
potentials.
No such point is expected to exist if the transition at $\mu_B = 0$ is first order.
No critical point has been found up to now in experiments with heavy 
ions, but the question is still open~\cite{tri1,tri2,tri3}.

\section{The method}

To establish the order of a phase transition using Monte Carlo simulations
is not an easy task. Since all quantities are analytical in a finite volume, one 
must study the finite size scaling (FSS) to look for developing singularities.
Here the FSS analysis is made intricate by the fact that, since simulations at zero quark
mass are not feasible, one has to consider both the finite spatial size $L_s$ and 
the finite quark mass $m_L=am_q$.
The FSS ansatz for the singular part of the free energy density $F_s$ around the chiral
critical point is:
\begin{equation}
F_s(\tau, m_L, L_s) = L_s^{-d} F_s\left(\tau L_s^{1/\nu}, m_L L_s^{y_h} \right) ,
\end{equation}
where $\tau$ is the reduced temperature $\tau=(1-T/T_c)$,
$\nu$ and $y_h$ are critical indexes. 
From the above equation we can derive the FSS of the specific heat:
\begin{equation}
C_V - C_0 \simeq  L_s^{\alpha/\nu} \phi_c \left(\tau L_s^{1/\nu}, m_L
L_s^{y_h} \right) , \label{scalcal}
\end{equation}
and of the susceptibility $\chi_m$ of the chiral condensate: 
\begin{equation}
\chi_m -\chi_0 \simeq L_s^{\gamma/\nu} \phi_\chi \left(\tau L_s^{1/\nu}, m_L L_s^{y_h} \right) . \label{scalord}
\end{equation}

In Ref.~\cite{nf2-I}\ we have approached the problem varying $m_L$ and $L_s$ in such a way that
the scaling variable $m_L L_s^{y_h}$ was fixed, thus reducing the FSS 
Eqs.~(\ref{scalcal}) and (\ref{scalord}), to one variable. 
Since the critical exponent $y_h$ must be fixed, one can only test a given universality class,
which in Ref.~\cite{nf2-I}\ was chosen to be $O(4)$. In that case no sign of scaling with the $O(4)$
critical exponents was found. In the present work we repeat the analysis fixing $y_h=3$ as expected 
for a first order transition.

\section{Simulation details}

As in Ref.~\cite{nf2-I}, the lattice discretization used is the standard staggered action.
The algorithm used in the new simulations for the test of first order scaling, is the RHMC.
All lattices have $L_t=4$.
New Monte Carlo data have been generated to match the $L_s=32$, $m_L=0.01335$ lattice 
used in Ref.~\cite{nf2-I}, thus fixing the parameter $m_L L_s^3$. New datasets for lattices 
with spatial extension $L_s=16,20,24$ have been produced.
For each of them ten different
values of $\beta$ spanning the entire critical region were
simulated with a total statistics of about 90k thermalized trajectories collected for
each of the lattices. 
The data collected is analized using a multi-histogram reweighting to extrapolate the observables to
intermediate couplings.
The backgrounds $C_0$ and $\chi_0$ appearing in Eqs.~(\ref{scalcal}) and (\ref{scalord}) are determined 
as in Ref.~\cite{nf2-I}, i.e. $C_0$ from a linear fit of the tails of the measured plaquette 
susceptibilities, and $\chi_0$ is fixed by a fit of the maxima of the chiral susceptibility in the 
critical region.

\begin{figure*}
\includegraphics*[height=.34\textwidth]{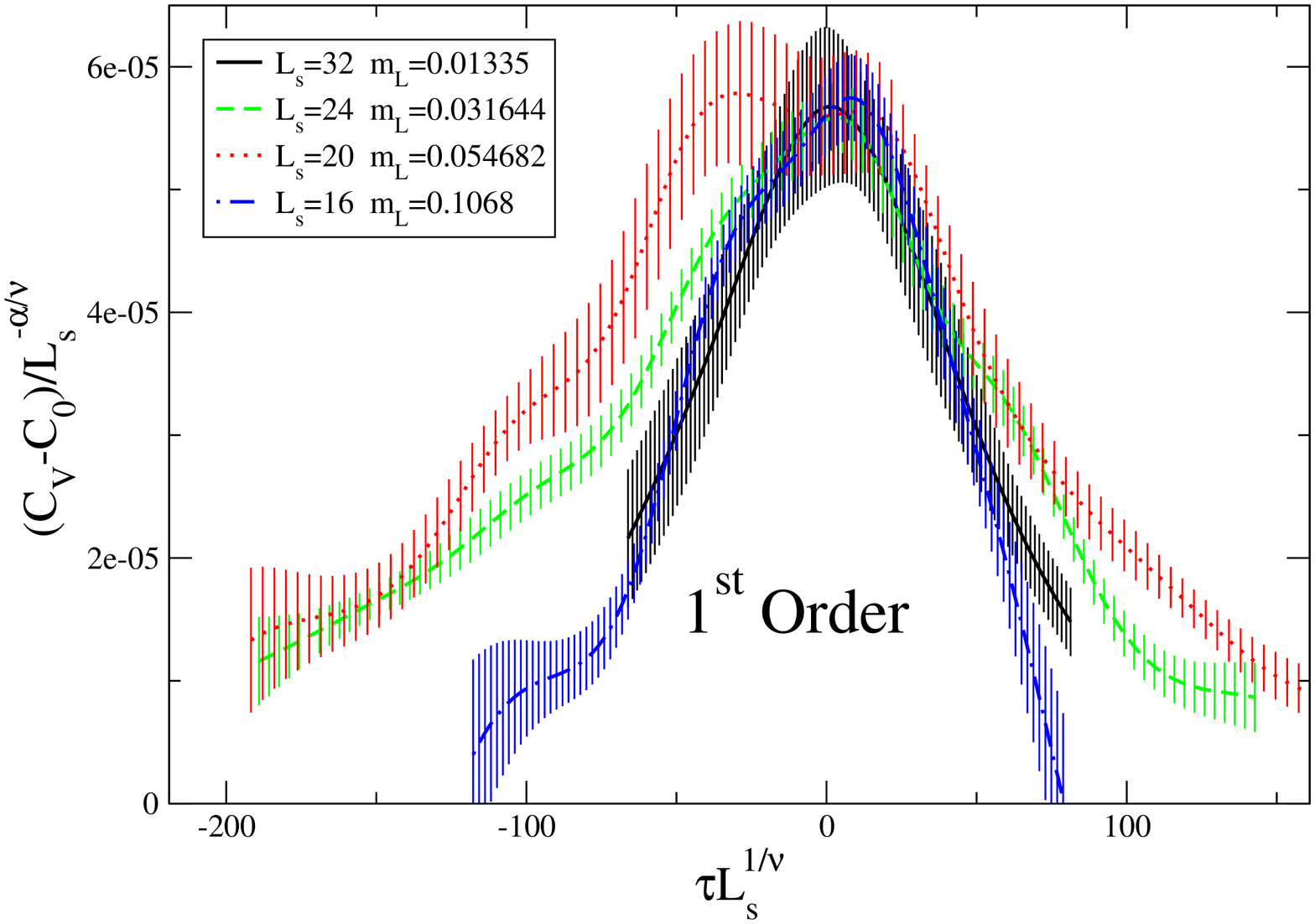}\hfill\includegraphics*[height=.34\textwidth]{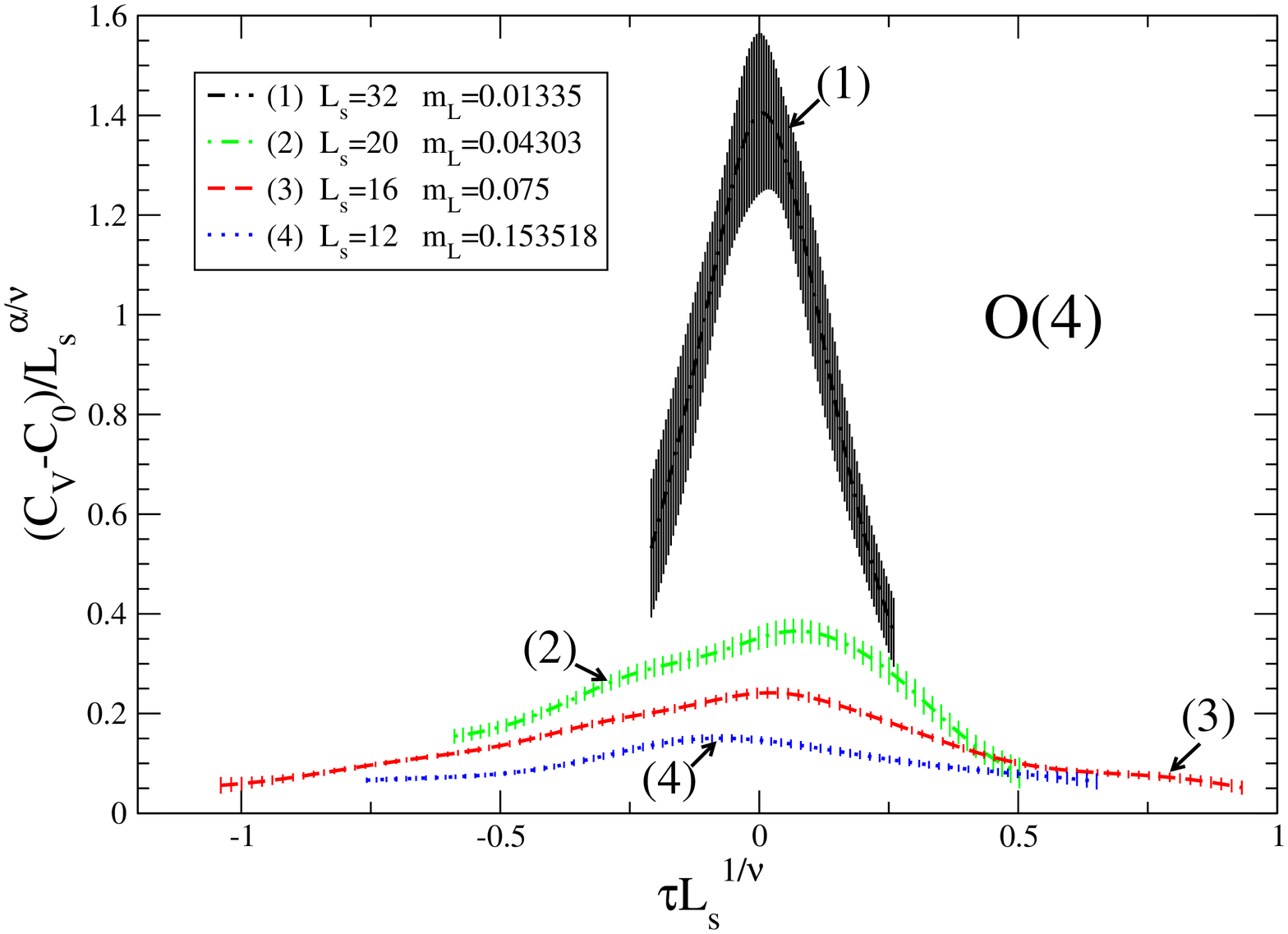}\\
\includegraphics*[height=.34\textwidth]{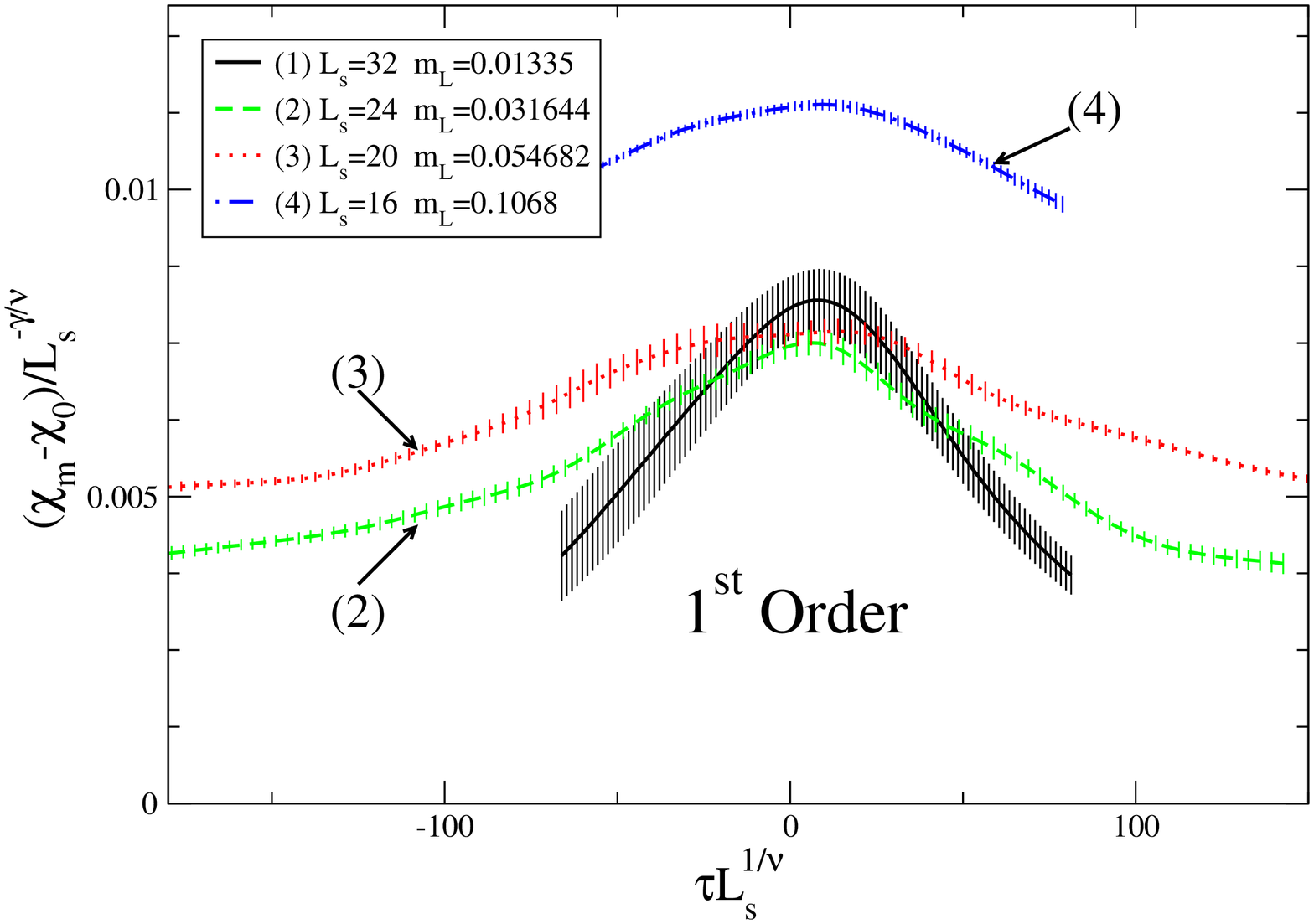}\hfill\includegraphics*[height=.34\textwidth]{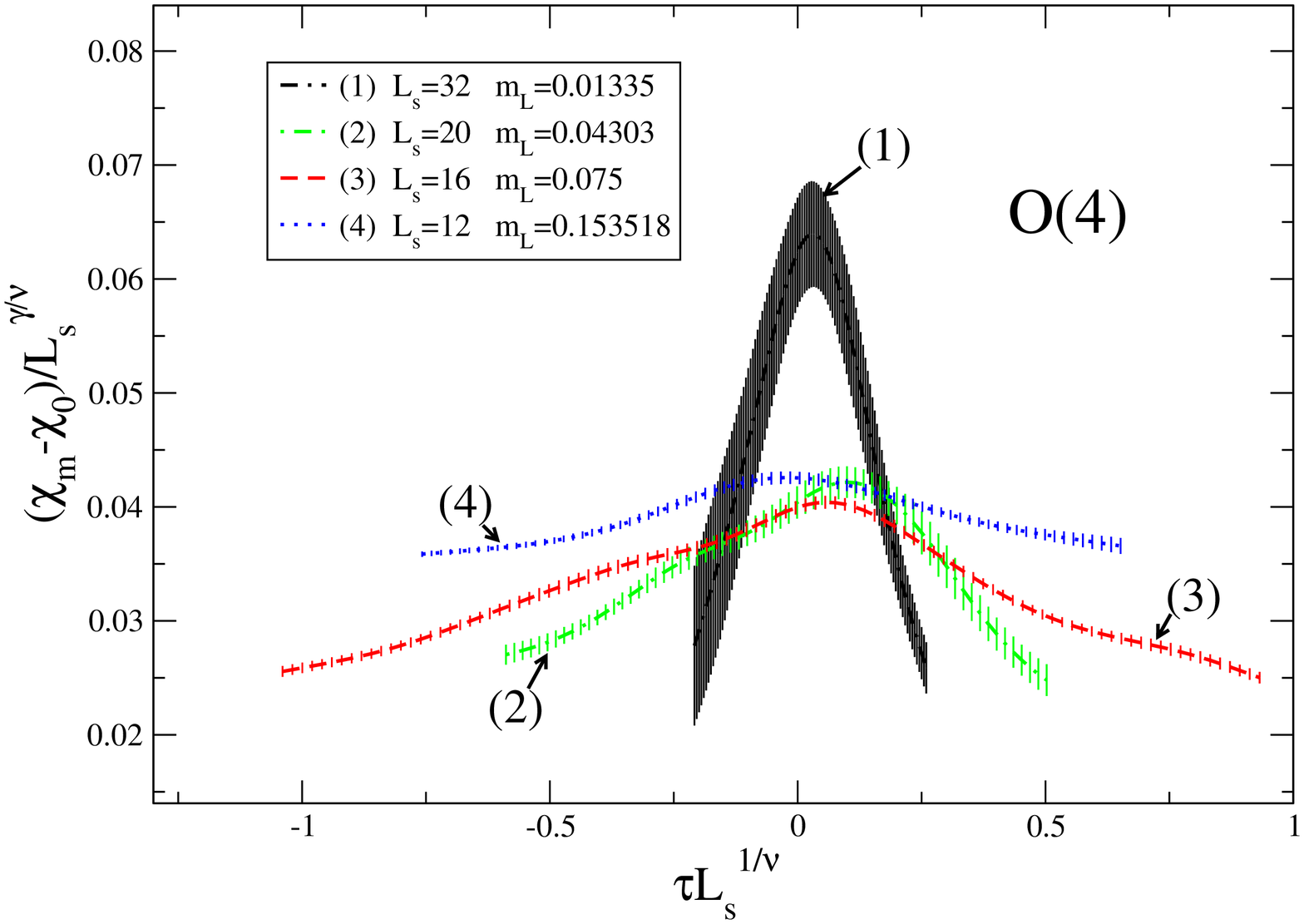}
\caption{Comparison of first order and $O(4)$ scaling of $C_V$ and $\chi_m$.\label{comparison}}
\end{figure*}

\begin{figure*}
\includegraphics*[height=.34\textwidth]{eqst1st}\hfill\includegraphics*[height=.34\textwidth]{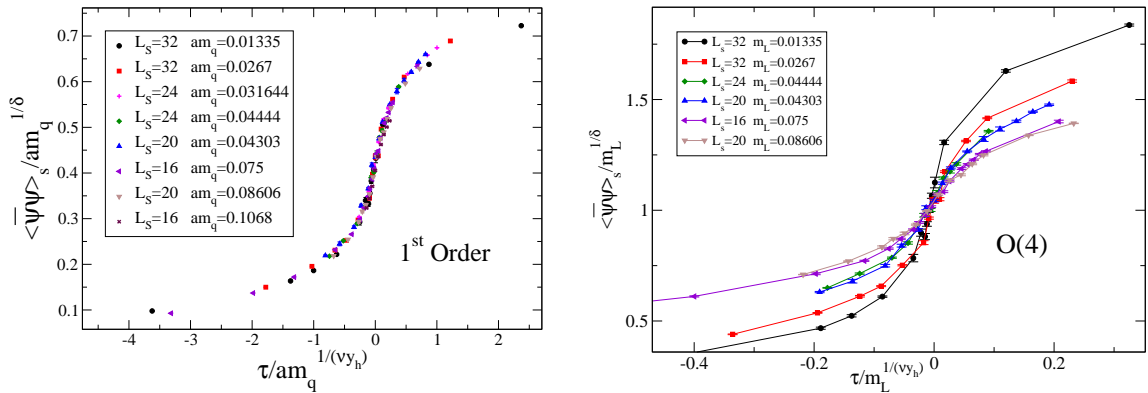}\\
\caption{Comparison of first order and $O(4)$ scaling of the subtracted chiral condensate.\label{eqst}}
\end{figure*}

\section{Test of scaling}

In Fig.~\ref{comparison} the scaling expected for a first order transition for the specific 
heat $C_V$ and of the chiral susceptibility $\chi_m$
is shown. The picture also includes the corrisponding figures for $O(4)$ from Ref.~\cite{nf2-I}.
For the first order case a reasonable scaling is observed for the specific heat $C_V$.
As remarked in Ref.~\cite{nf2-I}, $C_V$ is independent of any 
prejudice on the symmetry and on the order parameter.
Violations of the scaling Eq.~(\ref{scalord}) are observed for $\chi_m$
at larger values of the masses.
In fact Eq.~(\ref{scalord}) is expected to be valid for the susceptibility 
of the order parameter. At large masses chiral symmetry is badly broken 
and $\langle\bar\psi\psi\rangle$ is possibly not a good order parameter.
In our data, both of the present paper and of Ref.~\cite{nf2-I}, 
Eq.~(\ref{scalord}) seems to be violated for $m_L>0.05$.

The scaling of the chiral condensate (magnetic equation of state) can also be checked.
The comparison between first order and second order $O(4)$ is shown in Fig.~\ref{eqst}.
The first order scaling is very good. As in Ref.~\cite{nf2-I}, the scaling of the 
chiral condensate is obtained after a subtraction proportional to $m_L$ as $m_L\rightarrow 0$.
No sign of scaling is observed for the second order case.

\section{Conclusions}

The direct test of first order scaling
shows a good scaling for the specific heat and
for the chiral susceptibility at masses $m_L<0.05$.  
At larger masses scaling of the chiral susceptibility is broken, presumably because of strong breaking of 
the chiral symmetry.
By comparison with the similar test for $O(4)$, the first order scaling is clearly preferred over the 
second order. 
We can say that $O(4)$ is inconsistent with the lattice data while a first order transition 
describes well the observations. Possible effects due to the discretization must be taken into account. 
We believe that ultraviolect effects should 
be irrelevant with respect to the large volume behavior. However
the use of finer lattices and new simulation algorithms
to approach the chiral limit will possibly clarify this issue.

\section{Acknowledgments}
We acknowledge the APENEXT technical staff of INFN (Rome) where most
of the simulations were performed.

The work of C.P. has been supported in part by contract DE-AC02-98CH1-886
with the U.S. Department of Energy.

\end{document}